# Axisymmetric thermo-electro-elastic analysis of a piezoelectric half-space[1]

M.H. Kargarnovin, R. Hashemi, M. Hashemi and H. Sadeghi

School of Mechanical Engineering, Sharif University of Technology, Tehran, Iran

ABSTRACT

In this study, an analytical solution is presented for thermo-electro-elastic analysis of piezoelectric semi-infinite bodies. For this purpose, governing equations are derived for a transversely isotropic piezoelectric material under axisymmetric thermo-electro-mechanical loading condition. A general closed form analytical solution is presented for the complementary and particular parts of the components of displacement vector and also electric potential function. Then, boundary conditions are imposed and in that case an explicit solution is obtained for piezoelectric semi-infinite bodies. Results show that when a piezoelectric half-space is subjected to constant/ramp surface thermal loading the maximum absolute value of radial stress occurs at the surface of the body. Whereas, the maximum absolute value of stress in normal direction of the half-space surface occurs not on the surface but somewhere near the surface of the body. Moreover, the peak values for these stress curves in the case of combined loadings are closest to the half space surface and the farthest in the case of ramp decaying thermal loading. In contrast to the constant/ramp surface thermal loading no clear maximum point for electric field distribution for the case of combined loadings can be seen in the normal direction to the half-space boundary.

## 1. INTRODUCTION

Recently, piezoelectric materials have been used extensively as sensors and actuators in intelligent systems and structures, such as Micro-Electro-Mechanical-Systems (MEMS). Specific characteristics of piezoelectric materials, such as lightweight, small size and good frequency response, make them an ideal alternative to use in actuators and sensors. So, understanding mechanical behavior of piezoelectric materials under different loading conditions like thermal, mechanical or electrical loading is of significant importance. Numbers of different studies are available in which the effects of piezoelectricity of mechanical systems under thermo-electro-mechanical loading are analyzed [1-4].

---

[1] Submitted to *Mathematics and Mechanics of Solids*

In recent years, the electro-mechanical analysis of piezoelectric half-spaces has been of major concern because of its performance in technological applications. Consequently, many researchers have theoretically considered the analysis with this nature of loadings in the piezoelectric half-space medium. In this regard, Sosa and Castro [5] considered the problem of concentrated load acting on a piezoelectric half-plane. In this work state space formulation for electro-elasticity analysis of piezoelectric half-plane is presented and an implicit expression for the components of stresses and electric potential is obtained. Ding et al. [6] used Fourier transform to obtain a general solution for the problem of transversely isotropic piezoelectric medium. They used this formulation in the analysis of a piezoelectric half-plane subjected to concentrated forces. Rajapakse [7] used Fourier integral transform to find an analytical solution for the analysis of plane strain/stress of piezoelectric solids. In this regard, he considered the problem of half-space under line pressure loading and electric charge and obtained an analytical solution for components of stress tensor and electric field in the medium. Kuang et al. [8] considered the problem of collinear electrodes attached at the surface of a piezoelectric half-plane and presented an analytical solution for electro-elastic field in the half-plane. Liou and Sung [9] investigated an analytical solution for an anisotropic piezoelectric half-plane under surface electromechanical loading. They calculated the mechanical stress components (electric component) induced by electric loading (mechanical loading) on the boundary of the piezoelectric half-plane.

Since piezoelectric materials may be employed in thermal environment where temperature change is often one of the most important criteria in analysis of their behaviors, researchers tried to include this in the governing equations for studying thermopiezoelectricity [10-15]. In this regard, Ashida et al. [16, 17] and Tauchert et al. [18] proposed an analytical solution for axisymmetric piezothermoelastic of plates by a potential function method which uncouples governing equations.

Taking advantage of the benefits of the coupling between electric and mechanical fields in piezoelectric materials, piezoelectrics are used extensively as smart structures such as transducers, sensors and actuators. The optimal design of these smart structures requires the development of theoretical tools for the study of a variety of problems concerning the piezoelectric materials under different loading conditions. In the recent years, the application of transducers, sensors and actuators in high/low temperatures has attracted scientists to study the effect of thermal loading in addition to the effect of electro-mechanical loading on the behavior of piezoelectric materials [21-25]. Furthermore, the problem of piezoelectric half-space subjected to electro-mechanical loading is important both from theoretical and practical point of views and many researchers have studied this problem [26-32]. In view of these practical observations and also the lack of any theoretical studies on the coupled thermo-electro-mechanical problem of piezoelectric half-spaces has motivated the authors to seek for a closed form analytical solution for the analysis of piezoelectric half-spaces under coupled thermo-electro-mechanical loading.

However, as far as the authors know, to date no single work is reported on piezothermoelastic of half-spaces. Thus, this led us to consider the thermo-electro-elastic problem of a half-space under coupled thermo-electro-mechanical loading. In this paper, the governing equations for the analysis of a transversely isotropic piezoelectric half-space will be presented under axisymmetric conditions. Hankel transform will be used and an analytical form for the complementary and particular parts of the solution of the governing equations will be presented. Closed form expressions will be obtained for the stress components and electric displacements within the body. Then, a general solution will be presented for thermo-electro-elastic analysis of piezoelectric half spaces. In some worked out case studies, the effect of different loadings at the boundary surface of the half-space on the electro-elasticity of a piezoelectric half-space will investigated.

## 2. PROBLEM STATEMENT AND GOVERNING EQUATIONS

Consider a piezoelectric half-space, polarized along the $z$-direction, under coupled thermo-electro-elastic loading applied to the boundary, as shown in Figure 1. A cylindrical coordinates $(r,\theta,z)$ is assumed with the $z$-axis along the axis of symmetry of the cylinder. The constitutive equations for axisymmetric piezothermoelastic can be expressed as follows [16-18]:

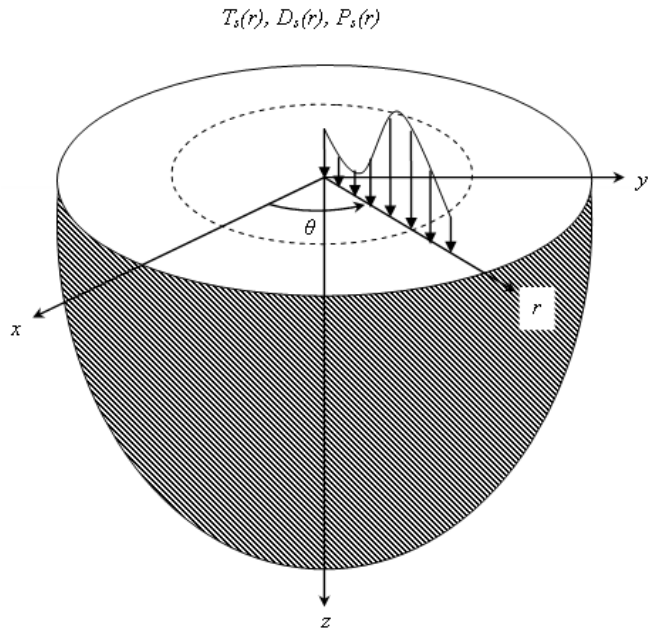

Fig. 1: A semi-infinite piezoelectric body under axisymmetric thermo-electro-mechanical loading.

$$\sigma_{rr} = c_{11}\varepsilon_{rr} + c_{12}\varepsilon_{\theta\theta} + c_{13}\varepsilon_{zz} - e_{31}E_z - \beta_1 \Delta T, \tag{1}$$

$$\sigma_{\theta\theta} = c_{12}\varepsilon_{rr} + c_{11}\varepsilon_{\theta\theta} + c_{13}\varepsilon_{zz} - e_{31}E_z - \beta_2 \Delta T, \tag{2}$$

$$\sigma_{zz} = c_{13}\varepsilon_{rr} + c_{13}\varepsilon_{\theta\theta} + c_{33}\varepsilon_{zz} - e_{33}E_z - \beta_3 \Delta T, \tag{3}$$

$$\sigma_{rz} = 2c_{44}\varepsilon_{rz} - e_{15}E_r, \tag{4}$$

$$D_r = 2e_{15}\varepsilon_{rz} + \epsilon_{11} E_r, \tag{5}$$

$$D_z = e_{31}\varepsilon_{rr} + e_{31}\varepsilon_{\theta\theta} + e_{33}\varepsilon_{zz} + \epsilon_{33} E_z - \beta_4 \Delta T, \tag{6}$$

where $\sigma_{ij}, \varepsilon_{ij}, D_{ij}, E_i, C_{ij}, e_{ij}$ and $\Delta T$ are components of stress tensor, strain tensor, electric displacement vector, electric field vector, elastic constants, piezoelectric constants and temperature change function (See Appendix A for details), respectively. Moreover, $\epsilon_{11}$ and $\epsilon_{33}$ are piezoelectric constants under zero or constant strain and $\beta_i (i=1,...,4)$ are known constants which can be expressed as follow

$$\beta_1 = c_{11}\alpha_r + c_{12}\alpha_\theta + c_{13}\alpha_z, \tag{7}$$

$$\beta_2 = c_{11}\alpha_\theta + c_{12}\alpha_r + c_{13}\alpha_z, \tag{8}$$

$$\beta_3 = c_{13}\alpha_r + c_{13}\alpha_\theta + c_{33}\alpha_z, \tag{9}$$

$$\beta_4 = e_{31}\alpha_r + e_{31}\alpha_\theta + e_{33}\alpha_z, \tag{10}$$

where $\alpha_i (i=r,\theta,z)$ are thermal expansion coefficients. It is known that in a transversely isotropic piezoelectric material, polarized in *z*-direction, thermal expansion coefficients in radial and tangential direction are the same, $\alpha_r = \alpha_\theta$ [19]. Consequently, in this case it is understood that $\beta_1 = \beta_2$.

The equilibrium equations and Gauss' equation representing the balance of electric displacements for a piezoelectric body under axisymmetric condition can be written as [20]

$$\frac{\partial \sigma_{rr}}{\partial r} + \frac{\partial \sigma_{rz}}{\partial z} + \frac{\sigma_{rr} - \sigma_{\theta\theta}}{r} = 0, \tag{11}$$

$$\frac{\partial \sigma_{rz}}{\partial r} + \frac{\partial \sigma_{zz}}{\partial z} + \frac{\sigma_{rz}}{r} = 0, \tag{12}$$

$$\frac{\partial D_r}{\partial r} + \frac{\partial D_z}{\partial z} + \frac{D_r}{r} = 0. \tag{13}$$

Furthermore, the strain-displacement relations and the relations between the electric field components, $E_i$, and electric potential $\varphi$ can be given by (Rajapakse 2005)

$$\varepsilon_{rr} = \frac{\partial u}{\partial r}, \quad \varepsilon_{\theta\theta} = \frac{u}{r}, \quad \varepsilon_{zz} = \frac{\partial w}{\partial z}, \quad \varepsilon_{rz} = \frac{1}{2}(\frac{\partial u}{\partial z} + \frac{\partial w}{\partial r}), \tag{14}$$

$$E_r = -\frac{\partial \varphi}{\partial r}, \quad E_z = -\frac{\partial \varphi}{\partial z}, \tag{15}$$

where $u$ and $w$ denote the components of displacement vector in $r$ and $z$ directions, respectively. Combination of equations (1)-(6) and equations (11)-(15) results in the following set of non-homogenous partially differential equations

$$\beta_1 \frac{\partial(\Delta T)}{\partial r} = c_{11}\left(\frac{\partial^2 u}{\partial r^2} + \frac{\partial}{\partial r}(\frac{u}{r}) + \frac{c_{44}}{c_{11}}\frac{\partial^2 u}{\partial z^2}\right) + \frac{\partial^2}{\partial r \partial z}[(c_{13}+c_{44})w + (e_{31}+e_{15})\varphi],$$

$$\beta_3 \frac{\partial(\Delta T)}{\partial z} = e_{15}\left(\frac{\partial^2 \varphi}{\partial r^2} + \frac{1}{r}\frac{\partial \varphi}{\partial r} + \frac{e_{33}}{e_{15}}\frac{\partial^2 \varphi}{\partial z^2}\right) + c_{44}\left(\frac{\partial^2 w}{\partial r^2} + \frac{1}{r}\frac{\partial w}{\partial r} + \frac{c_{33}}{c_{44}}\frac{\partial^2 w}{\partial z^2}\right) + (c_{44}+c_{13})\frac{\partial}{\partial z}\left(\frac{\partial u}{\partial r} + \frac{u}{r}\right), \quad (16)$$

$$\beta_4 \frac{\partial(\Delta T)}{\partial z} = -\in_{11}\left(\frac{\partial^2 \varphi}{\partial r^2} + \frac{1}{r}\frac{\partial \varphi}{\partial r} + \frac{\in_{33}}{\in_{11}}\frac{\partial^2 \varphi}{\partial z^2}\right) + e_{15}\left(\frac{\partial^2 w}{\partial r^2} + \frac{1}{r}\frac{\partial w}{\partial r} + \frac{e_{33}}{e_{15}}\frac{\partial^2 w}{\partial z^2}\right) + (e_{15}+e_{31})\frac{\partial}{\partial z}\left(\frac{\partial u}{\partial r} + \frac{u}{r}\right),$$

which are the governing equations for thermo-electro-elastic analysis of an axisymmetric semi-infinite piezoelectric body.

## 3. ANALYTICAL SOLUTION

*3.1 Solution Method*

In this section, a solution will be presented for the non-homogenous partial differential equations (16). The general solution of these equations can be expressed as the sum of the complementary and particular solutions. Herein, the complementary parts of the solution of the governing equations (16) are assumed to have the following forms

$$u_c = \int_0^\infty E(s) J_1(sr) e^{-ksz} ds, \qquad (17)$$

$$w_c = \int_0^\infty F(s) J_0(sr) e^{-ksz} ds, \qquad (18)$$

$$\varphi_c = \int_0^\infty G(s) J_0(sr) e^{-ksz} ds, \qquad (19)$$

where $E(s)$, $F(s)$ and $G(s)$ are arbitrary functions and $k$ is an arbitrary constant. Substituting equations (17)-(19) in the homogenous form of the governing equations (16), a set of algebraic equations will be found which can be written in the matrix form as

$$[P] \cdot \{E(s) \quad F(s) \quad H(s)\}^T = 0, \qquad (20)$$

where $[P]$ is a known matrix and can be given by

$$[P] = \begin{bmatrix} c_{44}k^2 - c_{11} & (c_{13}+c_{44})k & (e_{31}+e_{15})k \\ (c_{13}+c_{44})k & c_{44} - c_{33}k^2 & e_{15} - e_{33}k^2 \\ (e_{31}+e_{15})k & e_{15} - e_{33}k^2 & -\in_{11} + \in_{33} k^2 \end{bmatrix}. \qquad (21)$$

Subsequently, the characteristic equation of the homogenous form of the governing equations (16) can be expressed as $det([P])=0$ which can be written in the following form

$$k^6 + \Omega_1 k^4 + \Omega_2 k^2 + \Omega_3 = 0, \tag{22}$$

where $\Omega_i (i=1,..,3)$ are some known real constants and can be determined in terms of materials properties as follow

$$\Omega_1 = \frac{2c_{13}e_{15}e_{33} - 2e_{31}c_{33}e_{15} + 2c_{13}e_{33}e_{31} - c_{11}c_{33} \in_{33} + c_{13}^2 \in_{33}}{c_{44}c_{33} \in_{33} + c_{44}e_{33}^2}$$
$$+ \frac{2c_{44}e_{33}e_{31} - c_{33}e_{31}^2 - c_{33}e_{15}^2 - c_{44}c_{33} \in_{11} - c_{11}e_{33}^2 + 2c_{13}c_{44} \in_{33}}{c_{44}c_{33} \in_{33} + c_{44}e_{33}^2},$$

$$\Omega_2 = \frac{c_{11}c_{33} \in_{11} + c_{11}c_{44} \in_{33} - 2c_{13}e_{15}^2 - 2c_{13}e_{15}e_{31} + c_{44}e_{31}^2 + 2c_{11}e_{15}e_{33} - 2c_{13}c_{44} \in_{11} - c_{13}^2 \in_{11}}{c_{44}c_{33} \in_{33} + c_{44}e_{33}^2}, \tag{23}$$

$$\Omega_3 = -\frac{c_{11}c_{44} \in_{11} + c_{11}e_{15}^2}{c_{44}c_{33} \in_{33} + c_{44}e_{33}^2}.$$

Since $\Omega_i$ are real constants, the bi-cubic equation (22) has three pairs of roots $(\pm k_1, \pm k_2, \pm k_3)$, where $k_1$ is a positive real number and $k_2$ and $k_3$ are either positive real numbers or a pair of complex conjugates with positive real parts. Hence, the complementary parts of the governing equations (16) can be written as follows

$$u_c = \sum_{j=1}^{3} \int_0^\infty E_j(s) J_1(sr) e^{-k_j sz} ds, \tag{24}$$

$$w_c = \sum_{j=1}^{3} \int_0^\infty F_j(s) J_0(sr) e^{-k_j sz} ds, \tag{25}$$

$$\varphi_c = \sum_{j=1}^{3} \int_0^\infty G_j(s) J_0(sr) e^{-k_j sz}. \tag{26}$$

Furthermore, unknown coefficients $E_j(s)$, $F_j(s)$ and $G_j(s)$ are not independent and have some relations together and can be written in terms of an arbitrary function $H_j(s)$ as follows

$$E_j(s) = \delta_j H_j(s), \quad F_j(s) = \eta_j H_j(s), \quad G_j(s) = \xi_j H_j(s), \quad (j=1,2,3), \tag{27}$$

where $\delta_j$, $\eta_j$ and $\xi_j$ are known constants as

$$\delta_j = (c_{13} + c_{44})(e_{15} - e_{33}k_j^2)k_j - (c_{44} - c_{33}k_j^2)(e_{31} + e_{15})k_j, \tag{28}$$

$$\eta_j = (c_{13} + c_{44})(e_{15} + e_{31})k_j^2 - (c_{44}k_j^2 - c_{11})(e_{15} - e_{33}k_j^2), \tag{29}$$

$$\xi_j = (c_{44}k_j^2 - c_{11})(c_{44} - c_{33}k_j^2) - (c_{13} + c_{44})^2 k_j^2. \tag{30}$$

Three unknown functions $H_j(s)$ will be determined by imposing the mechanical and electrical boundary conditions.

To obtain a particular solution for the non-homogenous form of the governing equations (16), it is assumed that the components of displacement vector $u$, $w$ and electric potential function, $\varphi$ can be expressed in the following forms

$$u_p = \int_0^\infty B(s) J_1(sr) e^{-\sqrt{\frac{k_r}{k_z}} sz} ds, \tag{31}$$

$$w_p = \int_0^\infty C(s) J_0(sr) e^{-\sqrt{\frac{k_r}{k_z}} sz} ds, \tag{32}$$

$$\varphi_p = \int_0^\infty D(s) J_0(sr) e^{-\sqrt{\frac{k_r}{k_z}} sz} ds, \tag{33}$$

where $k_r$ and $k_z$ are heat conduction coefficients in $r$ and $z$ directions, respectively and $B(s)$, $C(s)$ and $D(s)$ are unknown functions which can be found by substituting equations (31)-(33) into the governing equations (16). This results in a set of algebraic equations which can be written in matrix form as $[M]\{B(s), C(s), D(s)\}^T = [N]$, where $[M]$ and $[N]$ are known matrices as following

$$[M] = \begin{bmatrix} -c_{11} + c_{44} \frac{k_r}{k_z} & (c_{13} + c_{44})\sqrt{\frac{k_r}{k_z}} & (e_{31} + e_{15})\sqrt{\frac{k_r}{k_z}} \\ (c_{13} + c_{44})\sqrt{\frac{k_r}{k_z}} & c_{44} - c_{33} \frac{k_r}{k_z} & e_{15} - e_{33} \frac{k_r}{k_z} \\ (e_{31} + e_{15})\sqrt{\frac{k_r}{k_z}} & e_{15} - e_{33} \frac{k_r}{k_z} & -\in_{11} + \in_{33} \frac{k_r}{k_z} \end{bmatrix}, \tag{34}$$

$$[N] = \frac{A(s)}{s} \begin{bmatrix} -\beta_1 \\ \beta_3 \sqrt{\frac{k_r}{k_z}} \\ \beta_4 \sqrt{\frac{k_r}{k_z}} \end{bmatrix}. \tag{35}$$

in which $A(s)$ is unknown function which can be determined by applying thermal boundary conditions (see Appendix A). Thus, unknown coefficient matrix $\{B(s), C(s), D(s)\}^T$ can be simply found as

$$\{B(s), C(s), D(s)\}^T = [M]^{-1}[N]. \tag{36}$$

Finally, the general solution of the governing equations (16) can be expressed as the sum of complementary and particular solutions as follows

$$u = \int_0^\infty J_1(sr)\left[ B(s)e^{-\sqrt{\frac{k_r}{k_z}}sz} + \sum_{j=1}^{3}\delta_j H_j(s)e^{-k_j sz}\right]ds, \tag{37}$$

$$w_p = \int_0^\infty J_0(sr)\left[ C(s)e^{-\sqrt{\frac{k_r}{k_z}}sz} + \sum_{j=1}^{3}\eta_j H_j(s)e^{-k_j sz}\right]ds, \tag{38}$$

$$\varphi_p = \int_0^\infty J_0(sr)\left[ D(s)e^{-\sqrt{\frac{k_r}{k_z}}sz} + \sum_{j=1}^{3}\xi_j H_j(s)e^{-k_j sz}\right]ds. \tag{39}$$

Using the constitutive equations (1)-(6), the stress components and electric displacements can be found as follow

$$\sigma_{rr} = \int_0^\infty \left[\left(R_0(s)J_0(sr)+R_1(s)\frac{J_1(sr)}{r}\right)e^{-\sqrt{\frac{k_r}{k_z}}sz} + \sum_{j=1}^{3}\left(\tilde{R}_{0j}(s)J_0(sr)+\tilde{R}_{1j}(s)\frac{J_1(sr)}{r}\right)e^{-k_j sz}\right]ds - \beta_1 \Delta T, \tag{40}$$

$$\sigma_{\theta\theta} = \int_0^\infty \left[\left(\Theta_0(s)J_0(sr)+\Theta_1(s)\frac{J_1(sr)}{r}\right)e^{-\sqrt{\frac{k_r}{k_z}}sz} + \sum_{j=1}^{3}\left(\tilde{\Theta}_{0j}(s)J_0(sr)+\tilde{\Theta}_{1j}(s)\frac{J_1(sr)}{r}\right)e^{-k_j sz}\right]ds - \beta_2 \Delta T, \tag{41}$$

$$\sigma_{zz} = \int_0^\infty J_0(sr)\left[Z_0(s)e^{-\sqrt{\frac{k_r}{k_z}}sz} + \sum_{j=1}^{3}\tilde{Z}_{0j}(s)e^{-k_j sz}\right]ds - \beta_3 \Delta T, \tag{42}$$

$$\sigma_{rz} = \int_0^\infty J_1(sr)\left[\Delta_1(s)e^{-\sqrt{\frac{k_r}{k_z}}sz} + \sum_{j=1}^{3}\tilde{\Delta}_{1j}(s)e^{-k_j sz}\right]ds, \tag{43}$$

$$D_r = \int_0^\infty J_1(sr)\left[\Gamma_1(s)e^{-\sqrt{\frac{k_r}{k_z}}sz} + \sum_{j=1}^{3}\tilde{\Gamma}_{1j}(s)e^{-k_j sz}\right]ds, \tag{44}$$

$$D_z = \int_0^\infty J_0(sr)\left[\Lambda_0(s)e^{-\sqrt{\frac{k_r}{k_z}}sz} + \sum_{j=1}^{3}\tilde{\Lambda}_{0j}(s)e^{-k_j sz}\right]ds - \beta_4 \Delta T, \tag{45}$$

where, $R_0(s)$, $R_1(s)$, $\tilde{R}_{0j}(s)$, $\tilde{R}_{1j}(s)$, $\Theta_0(s)$, $\Theta_1(s)$, $\tilde{\Theta}_{0j}(s)$, $\tilde{\Theta}_{1j}(s)$, $Z_0(s)$, $\tilde{Z}_{0j}$, $\Delta_1(s)$, $\tilde{\Delta}_{1j}(s)$, $\Gamma_1(s)$, $\tilde{\Gamma}_{1j}(s)$, $\Lambda_0(s)$ and $\tilde{\Lambda}_{0j}(s)$ are known functions and are given in Appendix B.

*3.2 Boundary conditions*

In this section, an axisymmetric piezoelectric half-space is considered and it is assumed that the half-space is subjected to axisymmetric thermal, electrical and mechanical loadings. Following boundary conditions are considered at $z=0$.

$$T = T_s(r), \quad D_z = D_s(r), \quad \sigma_{rz} = \tau_s(r), \quad \sigma_{zz} = P_s(r), \tag{46}$$

where $T_s(r)$, $D_s(r)$, $\tau_s(r)$ and $P_s(r)$ are the prescribed temperature distribution, electric displacement in $z$-direction, shear stress and normal pressure at the surface of the half-space, respectively. By substituting boundary conditions (46) in the stress components and electric displacements given in the equations (42), (43) and (45) a set of algebraic equation will result in. This set of equations can be written in the matrix form as $[X]\{H_1(s), H_2(s), H_3(s)\}^T = \{Y\}$, where $[X]$ and $\{Y\}$ are known matrices as follows

$$[X] = \begin{bmatrix} c_{13}\delta_1 - c_{33}\eta_1 k_1 - e_{33}\xi_1 k_1 & c_{13}\delta_2 - c_{33}\eta_2 k_2 - e_{33}\xi_2 k_2 & c_{13}\delta_3 - c_{33}\eta_3 k_3 - e_{33}\xi_3 k_3 \\ c_{44}\delta_1 k_1 + c_{44}\eta_1 + e_{15}\xi_1 & c_{44}\delta_2 k_2 + c_{44}\eta_2 + e_{15}\xi_2 & c_{44}\delta_3 k_3 + c_{44}\eta_3 + e_{15}\xi_3 \\ e_{31}\delta_1 - e_{33}\eta_1 k_1 + \in_{33} \xi_1 k_1 & e_{31}\delta_2 - e_{33}\eta_2 k_2 + \in_{33} \xi_2 k_2 & e_{31}\delta_3 - e_{33}\eta_3 k_3 + \in_{33} \xi_3 k_3 \end{bmatrix},$$

$$\{Y\} = \begin{Bmatrix} -c_{13}B(s) + c_{33}\sqrt{\dfrac{k_r}{k_z}}C(s) + e_{33}\sqrt{\dfrac{k_r}{k_z}}D(s) + \beta_3 \dfrac{A(s)}{s} + \overline{p}(s) \\ -c_{44}\sqrt{\dfrac{k_r}{k_z}}B(s) - c_{44}C(s) - e_{15}D(s) - \overline{\tau}(s) \\ -e_{31}B(s) + e_{33}\sqrt{\dfrac{k_r}{k_z}}C(s) - \in_{33}\sqrt{\dfrac{k_r}{k_z}}D(s) + \beta_4 \dfrac{A(s)}{s} + \overline{q}(s) \end{Bmatrix}, \tag{47}$$

in which $\overline{p}(s)$, $\overline{\tau}(s)$ and $\overline{q}(s)$ are

$$\overline{p}(s) = \int_0^\infty p(r) r J_0(sr) dr, \quad \overline{\tau}(s) = \int_0^\infty \tau(r) r J_1(sr) dr, \quad \overline{q}(s) = \int_0^\infty q(r) r J_0(sr) dr, \tag{48}$$

Hence, unknown coefficients matrix $\{H_1(s), H_2(s), H_3(s)\}^T$ can be simply found as

$$\{H_1(s), H_2(s), H_3(s)\}^T = [X]^{-1}\{Y\}. \tag{49}$$

## 4. NUMERICAL RESULTS

Based on obtained relations in previous section, a computer program is written using Maple software. After inputting temperature distribution function in its solver, the variation of stress components in $z$ and $r$ directions, as well as electric field in $z$-direction will be calculated. To study the effect of surface thermal loading and a combination of thermo-electro-mechanical loading on the behavior of a piezoelectric half-space, three different examples will be presented in this section. In all examples the used elastic and electrical material properties are given in Table 1 which corresponds to PZT-6B material [19]. Furthermore, the values of $k_i$'s for this material are calculated and presented in Table 2.

Table 1: Material properties of PZT-6B

| Elastic coefficients ($10^{10}$ N m$^{-2}$) | | | | | Piezoelectric coefficients (Cm$^{-2}$) | | | Dielectric coefficients ($10^{-10}$ Fm$^{-1}$) | | Thermal expansion coefficients ($10^{-6}$ K$^{-1}$) | | Thermal conductivity (W/m°C) | |
|---|---|---|---|---|---|---|---|---|---|---|---|---|---|
| $c_{11}$ | $c_{33}$ | $c_{12}$ | $c_{13}$ | $c_{44}$ | $e_{15}$ | $e_{31}$ | $e_{33}$ | $\epsilon_{11}$ | $\epsilon_{33}$ | $\alpha_z$ | $\alpha_r$ | $k_z$ | $k_r$ |
| 16.8 | 16.3 | 6.0 | 6.0 | 2.71 | 4.6 | 0.9 | 7.1 | 3.6 | 3.4 | 7 | 7 | 1.2 | 1.2 |

Table 2: Computed values of $k_i$ for PZT-6B.

| $k_i$ | | |
|---|---|---|
| 1 | 2 | 3 |
| 0.517598 | 1.01435 | 2.10143 |

*4.1 Piezoelectric half-space under constant surface thermal loading*

It is assumed that the prescribed temperature distribution has the following form at the surface of the half-space (see Figure 2)

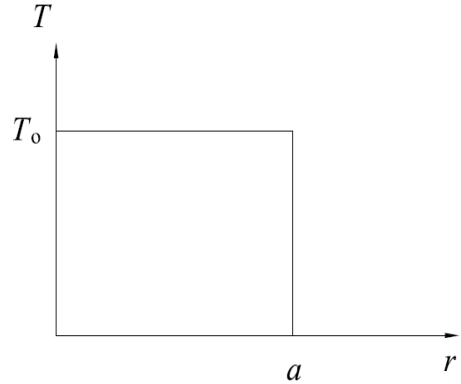

Fig. 2: Prescribed thermal boundary condition (constant temperature) at the surface of a piezoelectric half-space.

$$T_s(r) = T_i + T_0 H(a-r), \qquad (50)$$

where $H(a-r)$ and $a$ are; the Heaviside function and an arbitrary constant, respectively; and $T_i$ and $T_0$ are; initial temperature in the body and an arbitrary prescribed constant temperature at the surface of the half-space, respectively. The following values are used in our calculations

$$T_i = 300\,\text{K}, \quad T_0 = 100\,\text{K}, \quad a = 1\,\text{m}. \tag{51}$$

Figure 3 to Figure 5 show nondimensional stress distribution in *r* and *z* directions as well as distribution of non-dimensional electric field in *z*-direction vs. the *z*-direction for different values of $r/a$, respectively. It can be seen that for high values of $z/a$ the nondimensional radial stress in the body yields rapidly to a constant value no matter what the value of $r/a$ ratio would be. Also, it is shown that for an specific low value of $z/a$ the $\sigma_{rr}/\beta_3 T_0$ has smaller value for larger values of $r/a$. Referred to Figure 4 and Figure 5 it can be seen that for low values of $z/a$, $\sigma_{zz}/\beta_3 T_0$ and $e_{33} E_z / \beta_3 T_0$ primarily increase to a maximum level and then decrease and yield to a certain value no matter what the value of $r/a$ would be. Moreover, it is understood that $\sigma_{zz}/\beta_3 T_0$ curves attain their maximum values at a small distance under the surface. Whereas, $e_{33} E_z / \beta_3 T_0$ curves attain their maximum values at the surface of the body. Furthermore, it can be seen that the absolute maximum values of $\sigma_{zz}/\beta_3 T_0$ and $e_{33} E_z / \beta_3 T_0$ decrease with increasing $r/a$ ratio.

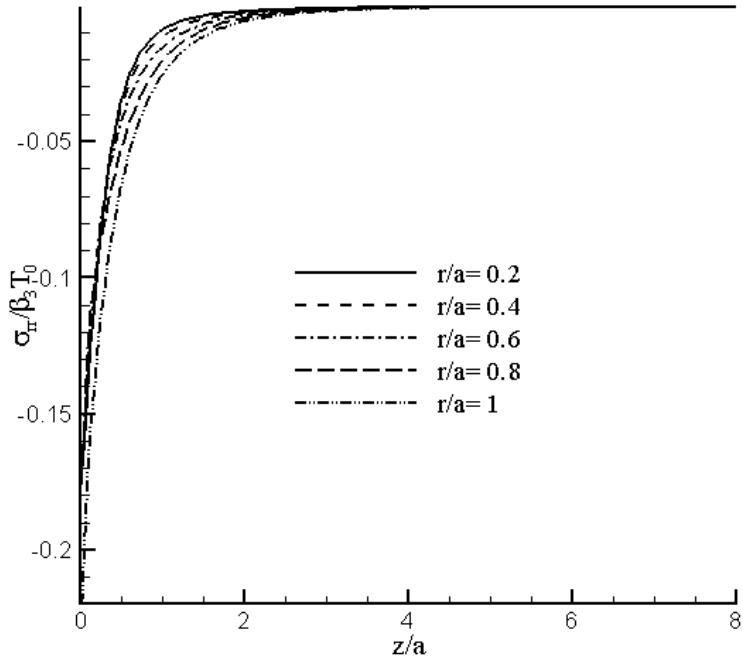

Fig. 3: Nondimensional stress distribution in *r*-direction in a piezoelectric half-space subjected to surface constant temperature loading.

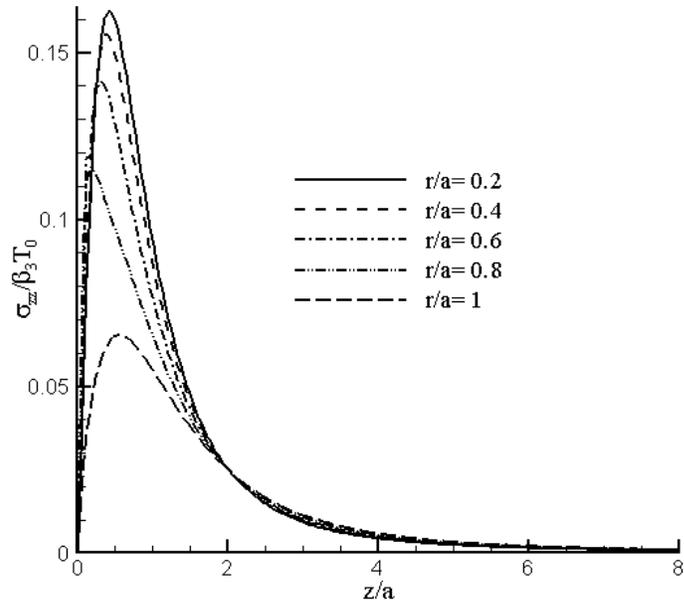

Fig. 4: Nondimensional stress distribution in *z*-direction in a piezoelectric half-space subjected to surface constant temperature loading.

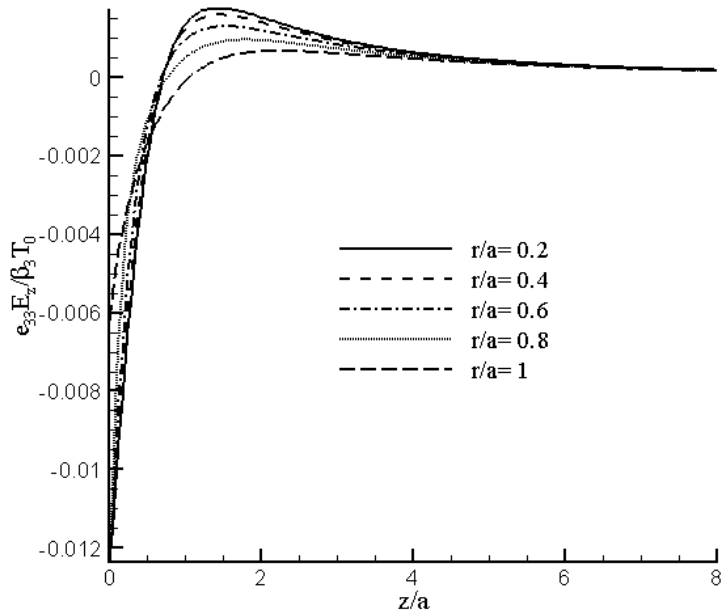

Fig. 5: Nondimensional electric field distribution in *z*-direction in a piezoelectric half-space subjected to surface constant temperature loading.

*4.2 Piezoelectric half-space under ramp decaying surface thermal loading*

In this section, it is assumed that the prescribed temperature distribution has the following form at the surface of the half-space (see Figure 6)

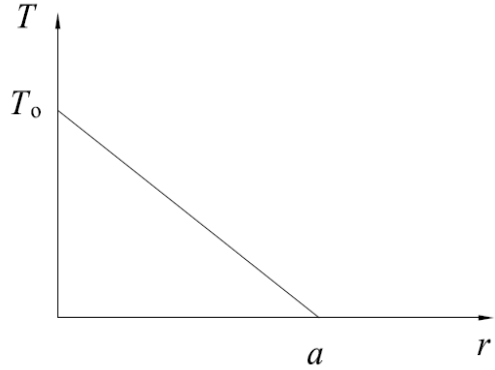

Fig. 6: Prescribed thermal boundary condition (ramp decaying temperature) at the surface of a piezoelectric half-space.

$$T_s(r) = T_i + T_0\left(1 - \frac{r}{a}\right) H(a-r), \tag{52}$$

where again $H(a-r)$ and $a$ are the Heaviside function and an arbitrary constant, respectively; and $T_i$ and $T_0$ are initial temperature in the body and an arbitrary prescribed constant temperature at the surface of the half-space, respectively. The following values are used in our calculations

$$T_i = 300 \text{ K}, \quad T_0 = 100 \text{ K}, \quad a = 1 \text{ m}. \tag{53}$$

Figure 7 to 9 show the nondimensional stress distribution in $r$ and $z$ directions as well as the distribution of nondimensional z-component of electric field vs. the $z$-direction for different values of $r/a$, respectively. It can be seen that for high values of $z/a$ the nondimenstional radial stress in the body yields rapidly to a constant value no matter what the value of $r/a$ ratio would be. Also, it is shown that the for an specific low value of $z/a$ the $\sigma_{rr}/\beta_3 T_0$ has larger value for smaller values of $r/a$. Referred to Figure 7 and 8 it can be seen that for low values of $z/a$, $\sigma_{zz}/\beta_3 T_0$ and $e_{33} E_z/\beta_3 T_0$ primarily increase to a maximum level and decrease and yield to a certain value no matter what the value of $r/a$ would be. Moreover, it is understood that $\sigma_{zz}/\beta_3 T_0$ curves attain their maximum values at a small distance under the surface. It is interesting to note that for the considered values of $r/a$, $e_{33} E_z/\beta_3 T_0$ curves attain their maximum values at the surface of the body for a range of $0.2 \le r/a \le 0.8$. On the other hand for $r/a = 1$ its maximum value occurs near the surface of the body. Furthermore, the absolute maximum values of $\sigma_{zz}/\beta_3 T_0$ and $e_{33}E_z/\beta_3 T_0$ decrease with increasing $r/a$ ratio.

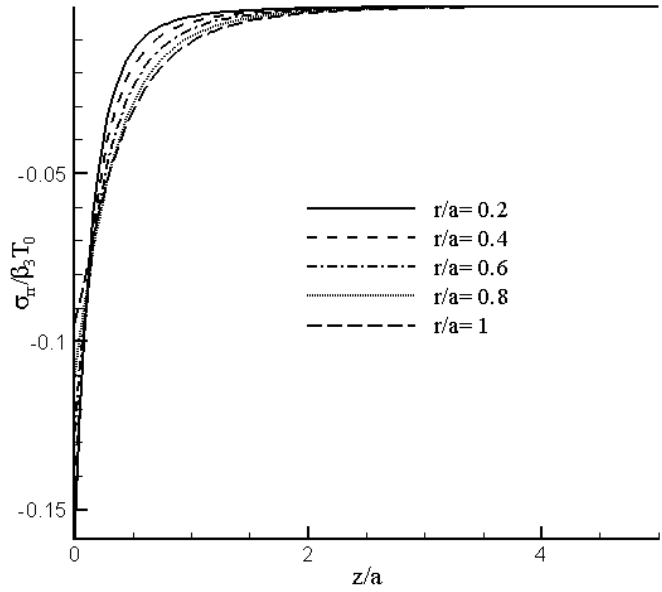

Fig. 7: Nondimensional stress distribution in *r*-direction in a piezoelectric half-space subjected to surface ramp temperature loading.

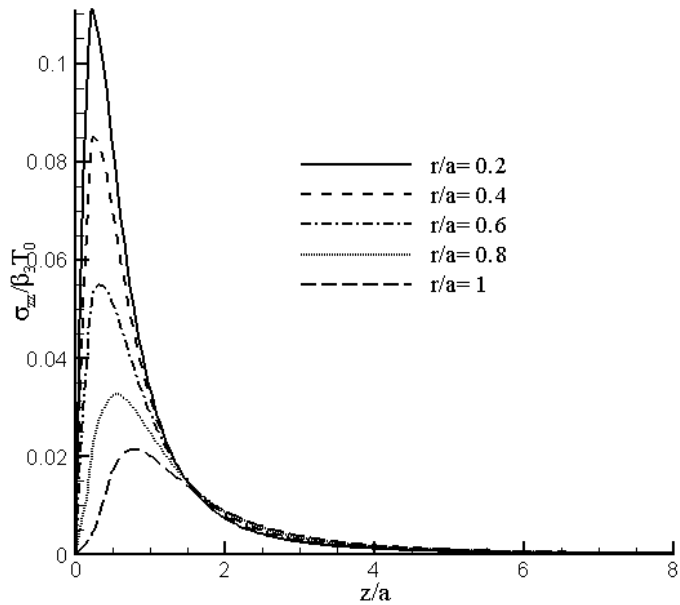

Fig. 8: Nondimensional stress distribution in *z*-direction in a piezoelectric half-space subjected to surface ramp temperature loading.

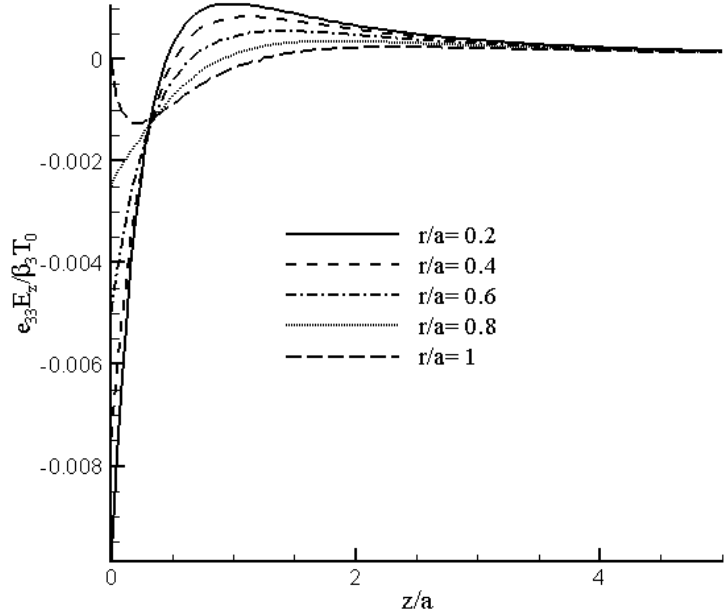

Fig. 9: Nondimensional electric field distribution in *z*-direction in a piezoelectric half-space subjected to surface ramp temperature loading.

*4.3 Piezoelectric half-space under constant surface thermo-electro-mechanical loading*

In this section, a combination of thermal, electrical and mechanical loadings in the following forms is imposed on the surface of this half-space.

$$T_s(r) = T_i + T_0 H(a-r), \tag{54}$$

$$q(r) = q_0 H(a-r), \tag{55}$$

$$p(r) = p_0 H(a-r), \tag{56}$$

where $q_0$ and $p_0$ denote magnitude of Heaviside function for applied electric charge and pressure, respectively and $T_0$, and $T_i$ are some constant temperatures defined in section 4.2. The following values are used in our calculations:

$$T_i = 300\,\text{K}, \quad T_0 = 100\,\text{K}, \quad q_0 = -0.01\,\text{Cm}^{-1}, \quad p_0 = -10\,\text{MPa}, \quad a = 1\,\text{m}. \tag{57}$$

By inputting above data into the developed computer program, stress distribution in *r* and *z* directions as well as the distribution of *z* component of electric field vs. the *z*-direction for different values of $r/a$ are shown in Figure 10 to Figure 12, respectively. It can be seen that for high values of $z/a$ the radial stress and electric field in *z*-direction in the body yields rapidly to a constant value no matter what the value of $r/a$ ratio would be, whereas stress in $z$-direction primarily increases to a maximum level and then decreases to a certain value. As it was observed in other cases, for this case, maximum value for $\sigma_z$ curves occurs also at a small distance under the surface of the half-space. It should be noted that for all values of $r/a$ the starting point in this Figure indicates the value of $\sigma_z$ at the surface of the half-space. It is interesting to see that in all Figures 10 to 12, the value of starting point for the curve of $r/a=1$ is not the same as four other curves. Even this behavior is more pronounced when one looks at the Figure 12. The reason for this type of trend certainly is linked to the characteristics of Heaviside function applied in the central region i.e. up to $r=a$. Note that $r=a$ is an interface between the region where the step load is applied ($r<a$) and load-less region ($r>a$) afterwards. By comparing the maximum value of $\sigma_z$ for three different loadings described in 4.1, 4.2 and this section, one can easily conclude that the peak value for this stress in the case of combined loadings is the closest to the half space surface and the farthest in the case of ramp decaying thermal loading. Similar comparison is made between these three types of loadings for $E_z$ distribution. Contrary to two other cases no clear maximum point can be seen in case of combined loadings near to the half-space boundary.

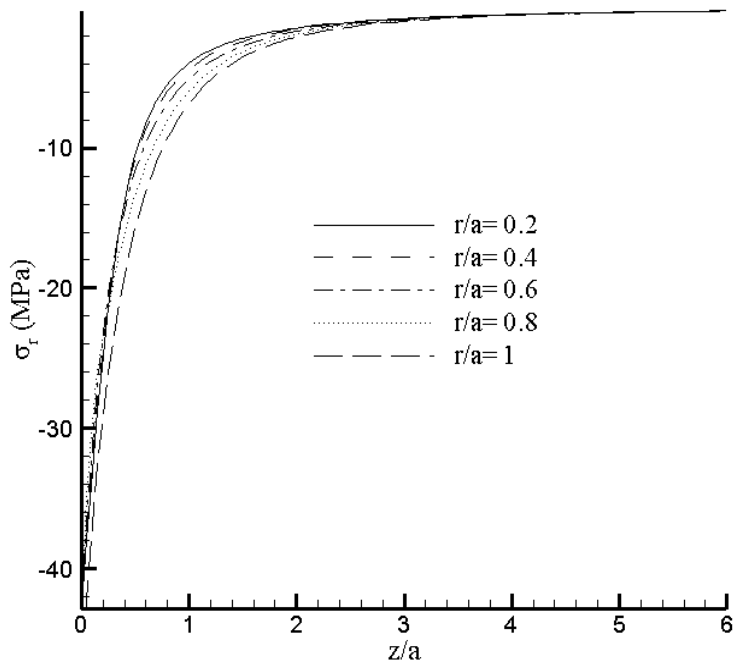

Fig. 10: Stress distribution (MPa) in r-direction in a piezoelectric half-space subjected to surface constant thermo-electro-mechanical loading.

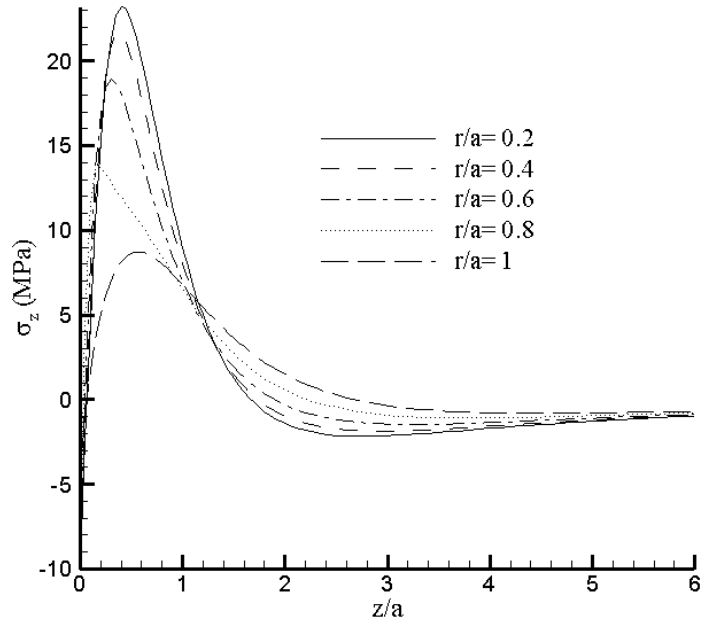

Fig. 11: Stress distribution (MPa) in z-direction in a piezoelectric half-space subjected to surface constant thermo-electro-mechanical loading.

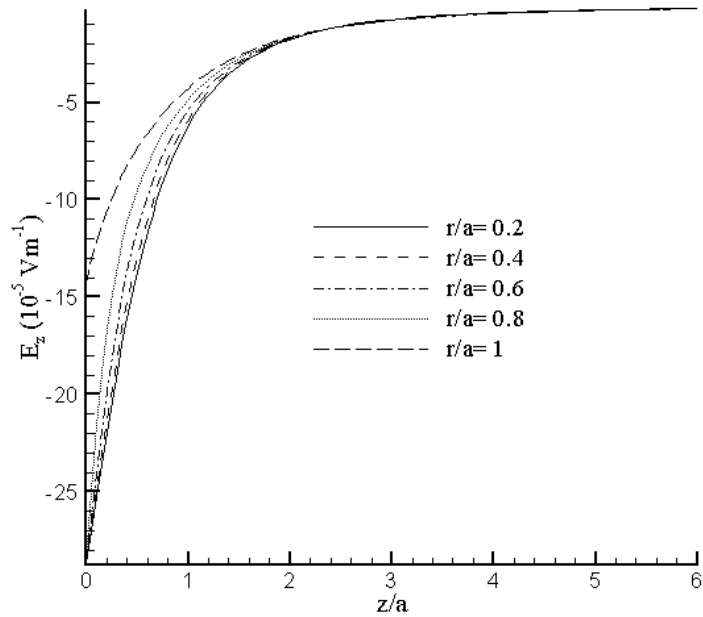

Fig. 12: Electric field distribution (V m$^{-1}$) in $z$-direction in a piezoelectric half-space subjected to surface constant thermo-electro-mechanical loading.

# 5. CONCLUSION

An analytical solution for thermo-electro-elastic analysis of transversely isotropic piezoelectric semi-infinite bodies under axisymmetric loading conditions is presented. The governing equations of transversely isotropic piezoelectric materials are derived under axisymmetric conditions. In this paper, it is assumed that the body is subjected to coupled thermo-electro-mechanical loading and an analytical solution is presented for stresses and electric field in a piezoelectric half-space. To study the effect of thermal loading on the thermo-electro-mechanical behavior of the piezoelectric half-space three different loading cases are analyzed. It is concluded that:

I. When a piezoelectric half-space is subjected to constant/ramp surface thermal loading the maximum absolute value of radial stress occurs at the surface of the body.
II. When a piezoelectric half-space is subjected to constant/ramp surface thermal loading the maximum absolute value of stress in *z*-direction occurs near the surface of the body.
III. Similar trend as stated in 1 and 2 of above is seen when one deals with combined thermo-electro-mechanical loading.
IV. The peak values in the $\sigma_z$ curves in the case of combined loadings are closest to the half space surface and the farthest in the case of ramp decaying thermal loading.
V. In contrast to the constant/ramp surface thermal loading no clear maximum point for $E_z$ distribution can be seen in case of combined loadings near to the half-space boundary.

**Appendix A:**

The steady state heat conduction equation for a transversely isotropic medium in axisymmetric condition can be written as

$$\frac{\partial^2 T(r,z)}{\partial r^2} + \frac{1}{r}\frac{\partial T(r,z)}{\partial r} + \frac{k_z}{k_r}\frac{\partial^2 T(r,z)}{\partial z^2} = 0, \tag{A.1}$$

where $k_r$ and $k_z$ are heat conduction coefficients in $r$ and $z$ directions, respectively; and $T(r,z)$ is temperature distribution in the medium. We define the temperature change function, $\Delta T(r,z)$, to be the difference between the temperature distribution in the body, $T(r,z)$, and initial temperature distribution in the body, $T_i(r,z)$, as follows

$$\Delta T(r,z) = T(r,z) - T_i(r,z). \tag{A.2}$$

It is assumed that the general solution of equation (A.1) can assumed to be in the following form

$$T(r,z) = A_0 + \int_0^\infty A(s) J_0(sr) e^{-\sqrt{\frac{k_r}{k_z}} sz} ds, \tag{A.3}$$

where $A_0$ and $A(s)$ are unknown functions which can be determined from thermal boundary conditions (A.4), and $J_0(sr)$ denotes the Bessel function of the first kind of order zero.

$$\begin{aligned} T &= T_i & \text{at} & \quad r, z \to \infty \\ T &= T_s(r) & \text{at} & \quad z = 0 \end{aligned} \tag{A.4}$$

Imposing the thermal boundary conditions (A.4), unknown coefficients in (A.3) can be determined as follows

$$A_0 = T_i, \tag{A.5}$$

$$A(s) = \int_0^\infty T_s(r) r J_0(sr) dr. \tag{A.6}$$

## Appendix B:

The definition of the functions used in description of stresses and electric displacements in equations (40)-(45) can be written as follow

$$R_0(s) = s\left(c_{11}B(s) - c_{13}\sqrt{\frac{k_r}{k_z}}C(s) - e_{31}\sqrt{\frac{k_r}{k_z}}D(s)\right), \quad R_1(s) = (c_{12} - c_{11})B(s),$$

$$\tilde{R}_{0j}(s) = s\left(c_{11}\delta_j - c_{13}\eta_j k_j - e_{31}\xi_j k_j\right)H_j(s), \quad \tilde{R}_{1j}(s) = (c_{12} - c_{11})\delta_j H_j(s),$$

$$\Theta_0(s) = s\left(c_{12}B(s) - c_{13}\sqrt{\frac{k_r}{k_z}}C(s) - e_{31}\sqrt{\frac{k_r}{k_z}}D(s)\right), \quad \Theta_1(s) = (c_{11} - c_{12})B(s),$$

$$\tilde{\Theta}_{0j}(s) = s\left(c_{12}\delta_j - c_{13}\eta_j k_j - e_{31}\xi_j k_j\right)H_j(s), \quad \tilde{\Theta}_{1j}(s) = (c_{11} - c_{12})\delta_j H_j(s),$$

$$Z_0(s) = s\left(c_{13}B(s) - c_{33}\sqrt{\frac{k_r}{k_z}}C(s) - e_{33}\sqrt{\frac{k_r}{k_z}}D(s)\right), \quad \tilde{Z}_{0j}(s) = s\left(c_{13}\delta_j - c_{33}\eta_j k_j - e_{33}\xi_j k_j\right)H_j(s), \qquad (B.1)$$

$$\Delta_1(s) = -s\left(c_{44}\sqrt{\frac{k_r}{k_z}}B(s) + c_{44}C(s) + e_{15}D(s)\right), \quad \Delta_{1j}(s) = -s\left(c_{44}\delta_j k_j + c_{44}\eta_j + e_{15}\xi_j\right)H_j(s),$$

$$\Gamma_1(s) = -s\left(e_{15}\sqrt{\frac{k_r}{k_z}}B(s) + e_{15}C(s) - \in_{11} D(s)\right), \quad \tilde{\Gamma}_{1j}(s) = -s\left(e_{15}\delta_j k_j + e_{15}\eta_j - \in_{11}\xi_j\right)H_j(s),$$

$$\Lambda_0(s) = s\left(e_{31}B(s) - e_{33}\sqrt{\frac{k_r}{k_z}}C(s) + \in_{33}\sqrt{\frac{k_r}{k_z}}D(s)\right), \quad \tilde{\Lambda}_{0j}(s) = s\left(e_{31}\delta_j - e_{33}\eta_j k_j + \in_{33}\xi_j k_j\right)H_j(s).$$